\documentclass[11pt, a4paper]{article}
\pdfoutput=1
\usepackage{jcappub}  
\usepackage{savesym}
\usepackage{graphicx}
\usepackage{subfigure}
  \expandafter\let\csname equation*\endcsname\relax
  \expandafter\let\csname endequation*\endcsname\relax 
\usepackage{amsmath}
\usepackage{verbatim}
\usepackage{array}
\newcommand{\be}{\begin{eqnarray}}
\newcommand{\beq}{\begin{equation}}
\newcommand{\bee}{\begin{enumerate}}
\newcommand{\bit}{\begin{itemize}}

\def\CC{cosmological constant }

\def\bkt#1{\left(#1\right)} 

\def\ee{\end{eqnarray}}
\def\eeq{\end{equation}}
\newcommand{\eee}{\end{enumerate}}

\newcommand{\eit}{\end{itemize}}

\def\ff{\phantom{.}}

\newcommand{\ii}{\textit}

\def\lab{\label}

\newcommand{\mpl}{m_{\mbox{\scriptsize{Pl}}}}
\newcommand{\mb}{\mathbf}

\def\pr{\prime}
\def\re#1{(\ref{#1})}

\def\sub#1{_{\mbox{\scriptsize{#1}}}}

\title{A gauge-invariant approach to interactions in the dark sector}

\author{William J. Potter and Sirichai Chongchitnan}

\affiliation{Oxford Astrophysics, \\
 Denys Wilkinson Building, Keble Road, Oxford, OX1 3RH, United Kingdom}
\emailAdd{Will.Potter@astro.ox.ac.uk}
\emailAdd{siri@astro.ox.ac.uk}
\abstract{We outline a gauge-invariant framework to calculate cosmological perturbations in dark energy models consisting of a scalar field interacting with dark matter via energy and momentum exchanges. Focusing on three well-known models of quintessence and three common types of dark sector interactions, we calculate the matter and dark energy power spectra as well as the Integrated Sachs-Wolfe (ISW) effect in these models. We show how the presence of dark sector interactions can produce a large-scale enhancement in the matter power spectrum and a boost in the low multipoles of the cosmic microwave background anisotropies. Nevertheless, we find these enhancements to be much more subtle than those found by previous authors who model dark energy using simple ansatz for the equation of state. We also address issues of instabilities and emphasise the importance of momentum exchanges in the dark sector.} 


\begin{document}
\maketitle
\flushbottom

\section{Introduction}


Observation of type Ia supernovae at high redshift, large scale structure surveys and observations of the cosmic microwave background (CMB) all indicate that the Universe is currently undergoing accelerated expansion \cite{1998AJ....116.1009R, 2009ApJS..180..330K, 2008ApJ...686..749K, 2009ApJS..182..543A}.   The most convincing explanation for this phenomenon is that the Universe is filled in great abundance with a form of energy with negative pressure. This so-called `dark energy' (DE) can lead to an accelerated cosmic expansion if its equation of state $w$ satisfies $w<-1/3$, although observations currently constrain $w$ to be close to $-1$. At present, the cosmological constant model in which $w$ is exactly $-1$ appears to be the most robust model for dark energy \cite{2009ApJS..180..330K, 2001sfu..conf..179E}.

However, there remain unsolved problems with the physical origin and the exact value of the cosmological constant. It is well known that the value of the cosmological constant, interpreted as vacuum energy arising from particle-antiparticle annihilation, and that consistent with observation can differ by as much as 120 orders of magnitude \cite{2007NuPhS.173....6D}. Moreover, the value of the cosmological constant, however miniscule, must be set at its precise value at the initial inflationary epoch, with the slightest deviation from this value leading to disastrous consequences \cite{1992ARA&A..30..499C}. 


Over a decade since the discovery of cosmic acceleration, there are now an overwhelming number of dark energy alternatives to the cosmological constant (see e.g. \cite{2006IJMPD..15.1753C} for a review). One of the earliest and the simplest of such models is that of a single light scalar-field model of dynamical dark energy dubbed `quintessence' \cite{1988PhRvD..37.3406R,1988NuPhB.302..668W, 2002PhRvD..66b3516D}. Unlike the cosmological constant, many quintessence models exhibit an attractor behaviour and thus do not require initial conditions that are extremely fine-tuned. The late-time acceleration can also be dynamically `switched on' without requiring a cosmic coincidence \cite{2002PhRvD..66d3523S}.


Since quintessence can be interpreted as a field with non-zero mass, it is possible that dark energy could cluster like dark matter, or even interact with dark matter. The possibility that there may be interactions in the dark sector has previously been explored by several authors  \cite{PhysRevD.68.023514, 2008JCAP...07..020V, 2010MNRAS.402.2355V, 2011arXiv1103.0694T, PhysRevD.72.043516, PhysRevD.74.043521, PhysRevD.78.123514, PhysRevD.77.103003, PhysRevD.78.083538, 2008JCAP...04..007L, PhysRevD.80.063530, 2009PhRvD..79d3526J, 2007JCAP...06..020M, 2009ApJ...697.1946V, 2009PhRvD..79d3522C, 1995A&A...301..321W}. It is common in the previous works to either neglect dark energy perturbations or model dark energy using some phenomenological form of $w$ (e.g. $w=$ constant). These simplistic approaches may not accurately reproduce the dynamics of quintessence models as suggested by \cite{2009PhRvL.103o1303P, 2010MNRAS.401.2181J}. 


The primary aim of this paper is to investigate the prospects of detecting the observational signatures of dark sector interactions. We shall consider three well-known models of quintessence and study the effects of dark sector interactions in both the background and in the linear perturbations. In particular, we shall examine the effects of non-zero interactions on the linear matter power spectrum and the integrated Sachs-Wolfe (ISW) using the gauge-invariant approaches of Kodama and Sasaki \cite{1984PThPS..78....1K}. Unlike previous works on this subject, our approach takes into account dark energy perturbations, including the perturbative effects of energy and momentum transfer in the dark sector. Issues of instabilities previously discovered in \cite{2008JCAP...07..020V, 2010MNRAS.402.2355V, 2009PhRvD..79d3522C} will also be addressed.

\section{Quintessence models}\lab{models}

Quintessence refers to a scalar field, $\phi$, evolving along a potential, $V(\phi)$, which becomes sufficiently flat at late times, leading to cosmic acceleration (see e.g. \cite{2006IJMPD..15.1753C}). At early times, the dark energy contribution to the overall energy density is small \cite{1998PhRvD..58b3503F}.  Some classes of potentials exhibit tracking behaviour whereby the dark energy density tracks below that of matter and radiation and only comes to dominate the universe at late times \cite{1998PhRvD..57.4686C}.  Another interesting feature of quintessence is the existence of attractor solutions \cite{1998PhRvD..58b3503F} which greatly reduce the need for the initial values of the dark energy parameters to be finely tuned. 

We will focus our analysis on three models of quintessence, namely, the Ratra-Peebles, SUGRA and double exponential potentials summarised below. From this point on, all equations will be in natural units with $c=\hbar=1$. We also write $\bar{\kappa}^{2}=8\pi G$. 

\subsection{Ratra-Peebles}

Ratra and Peebles (RP) \cite{1988PhRvD..37.3406R} showed that the class of inverse power law potentials of the form

\begin{equation}
V(\phi)=\frac{M^{4+\alpha}}{\phi^{\alpha}}, \,\,\, \alpha>0,
\end{equation}
tracks the equation of state during matter and radiation-dominated eras, and dark energy only becomes dominant at late times. This behaviour holds for a wide range of initial conditions.  We will use $M=4.1\times10^{-28} \mpl$ and $\alpha=0.5$ as indicated by \cite{2010MNRAS.401..775A} who obtained these values from a likelihood analysis of the WMAP 5-year data \cite{2009ApJS..180..330K} and the `union' supernovae data \cite{2008ApJ...686..749K}.

\subsection{SUGRA}

The SUGRA model with potential

\begin{equation}\label{SU}
V(\phi)=\frac{M^{4+\alpha}\exp\bkt{\frac{1}{2}\bar{\kappa}^{2}\phi^{2}}}{\phi^{\alpha}},
\end{equation}
is motivated by supersymmetry and differs from the Ratra-Peebles (RP) potential by an exponential `supergravity' correction factor \cite{1999PhLB..468...40B}. This factor can be shown to push the equation of state closer to $-1$ at late times.  We will use the values $M=1.7\times10^{-25} \mpl$ and $\alpha=1$ as indicated by  \cite{2010MNRAS.401..775A}.  Again, the SUGRA potential has attractor solutions and is viable for a wide range of initial conditions.

\subsection{Double exponential}

The double exponential (DExp) potential 

\begin{equation}
V(\phi)= M_{1}e^{-\lambda_{1}\phi} + M_{2}e^{-\lambda_{2}\phi},
\end{equation}
allows dark energy to track the equation of state of radiation and matter and leads to acceleration at late times  \cite{2000PhRvD..61l7301B}.  We shall use $M_{1}=10^{-14}\mpl$, $M_{2}=10^{-13}\mpl$, $\lambda_{1}=9.43$ and $\lambda_{2}=1$ which, as shown in  \cite{2008JCAP...07..007B}, should give rise to observables that are significantly different from the \CC but are still broadly consistent with observational constraints.


\section{Evolution of background energy densities}

We assume an isotropic, homogeneous and spatially flat background as described by the Friedmann-Robertson-Walker (FRW) metric,
\begin{equation}\label{FRW}
ds^{2}=-dt^{2}+a(t)^{2}dx^{i}dx_{i},
\end{equation}
where $a(t)$ is the scale factor as a function of cosmic time, $t$, and summation is implied over $i=1,2,3$. We assume that the background energy density comprises matter, radiation and dark energy (denoted by subcripts $m$, $r$ and $\phi$ respectively). The total energy density, $\rho=\rho_m+\rho_r+\rho_\phi$, satisfies the Friedmann and acceleration equations
\begin{equation}
\label{H0}
H^{2}\equiv\bkt{\dot{a}\over a}^2=\frac{\bar{\kappa}^{2}}{3}(\rho_{m}+\rho_{r}+\rho_{\phi}),
\end{equation}
\begin{equation}
\label{Ac2}
\dot{H}=-\frac{\bar{\kappa}^{2}}{2}\left(\rho_{m}+\frac{4}{3}\rho_{r}+\rho_{\phi}+p_{\phi}\right),
\end{equation}
as well as the energy conservation equation
\beq
\label{Fl}
\dot{\rho}+3H(\rho+p)=0.
\eeq
Here overdots are derivatives with respect to $t$ and $p_x$ is the pressure of component $x$ related to its energy density via the equation of state $w_x=p_x/\rho_x$. In the above, we have used  $w_{m}=0$, $w_{r}=1/3$ and $w_{\phi}$ defined below.

The quintessence field evolves via the Klein-Gordan equation
\begin{equation}\label{KG}
\ddot{\phi}+3H\dot{\phi}+V^{\prime}(\phi)=0.
\end{equation}
Its energy density, pressure and equation of state are given by
\begin{equation}
\rho_\phi={\dot\phi^2 \over 2}+V(\phi), \quad p_\phi ={\dot\phi^2\over2}-V(\phi),
\end{equation}
\begin{equation}\label{wp}w_{\phi}=\frac{p_{\phi}}{\rho_{\phi}}=\frac{\dot\phi^2/2-V(\phi)}{\dot\phi^2/2+V(\phi)}.
\end{equation}
The case in which $\dot\phi=0$ reduces to the cosmological constant with $w_\Lambda=-1$. Note that at present $w_\phi$ is observationally constrained at low redshift to a value close to $w_{\phi}=-1$ with $\sim10\%$ accuracy \cite{2011ApJS..192...18K,2011ApJ...730..119R}. 

We model interactions between dark matter and dark energy by including a source term, $Q_{x}$, in the evolution equations for the dark energy and dark matter background energy densities.  
\begin{equation}
\dot{\rho_{x}}=-3H\rho_{x}(1+w_{x})+Q_{x}.
\end{equation}
Non-zero $Q$ thus represents the energy flow from one component to another. To conserve the total energy, we require
\begin{equation}
\sum_{x}{Q}_{x}=0.\label{Q5}
\end{equation}
If interactions are purely between the dark matter and dark energy (i.e. interactions only in the dark sector) then
 \begin{equation}
Q_{m}=-Q_{\phi}.\label{nor}
\end{equation}

\subsection{Evolution equations}
To evolve the background energy densities with time, we use the set of dimensionless `energy' variables defined in \cite{1998PhRvD..57.4686C}, 
\begin{center}
$x=\dfrac{\bar{\kappa}\dot{\phi}}{\sqrt{6}H}$, \,\,\,\,\,\,     
$y=\dfrac{\bar{\kappa}}{H}\sqrt{\dfrac{V}{3}}$,\\
\begin{equation}\label{Dv}
r=\dfrac{\bar{\kappa}}{H}\sqrt{\dfrac{\rho_{r}}{3}}, \,\,\,\,\,\, 
m=\dfrac{\bar{\kappa}}{H}\sqrt{\dfrac{\rho_{m}}{3}}, \,\,\,\,\,\,  
\lambda=-\dfrac{1}{\bar{\kappa}V}\dfrac{dV}{d\phi}.
\end{equation}
\end{center}
Intuitively, $x,y,r$ and $m$ can be thought of as (the square-root of) the ratio of energy density of each component to the total energy density ($\sim3H^2$).  
Converting the time variable in Equations$\nolinebreak \,\ref{Ac2},\, \ref{Fl}$ and $\ref{KG}$ to the `$e$-fold' number, ${N=\ln a}$, we obtain

\setlength\arraycolsep{0.1em}
\begin{eqnarray}
\dfrac{d\ln H}{dN}&=&-\frac{3}{2}\left(1+x^{2}-y^{2}+\frac{r^{2}}{3}\right),\label{E1}\\
\frac{dx}{dN}&=&\sqrt{\frac{3}{2}} \lambda y^{2}-x\left(3+\frac{d\ln H}{dN}\right)+\frac{\overline{\kappa}^{2}}{6H^{3}x}Q_{\phi},\label{Q16}\\
\frac{dy}{dN}&=&-\sqrt{\frac{3}{2}}\lambda xy-y\frac{d\ln H}{dN},\\
\frac{dr}{dN}&=&-r\left(2+\frac{d\ln H}{dN}\right),\\
\frac{dm}{dN}&=&-m\left(\frac{3}{2}+\frac{d\ln H}{dN}\right)+\frac{\overline{\kappa}^{2}}{6H^{3}m}Q_{m}.
\label{E5}
\end{eqnarray}
These variables obey the Friedmann constraint derived from dividing Equation$\nolinebreak \,\ref{H0}$ by $H$.
\begin {equation}\label{Nm}
x^{2}+y^{2}+r^{2}+m^{2}=1, 
\end {equation}


\begin{figure}
\centering
\includegraphics[width=6.5cm,clip=true, trim=1cm 1cm 1cm 0cm]{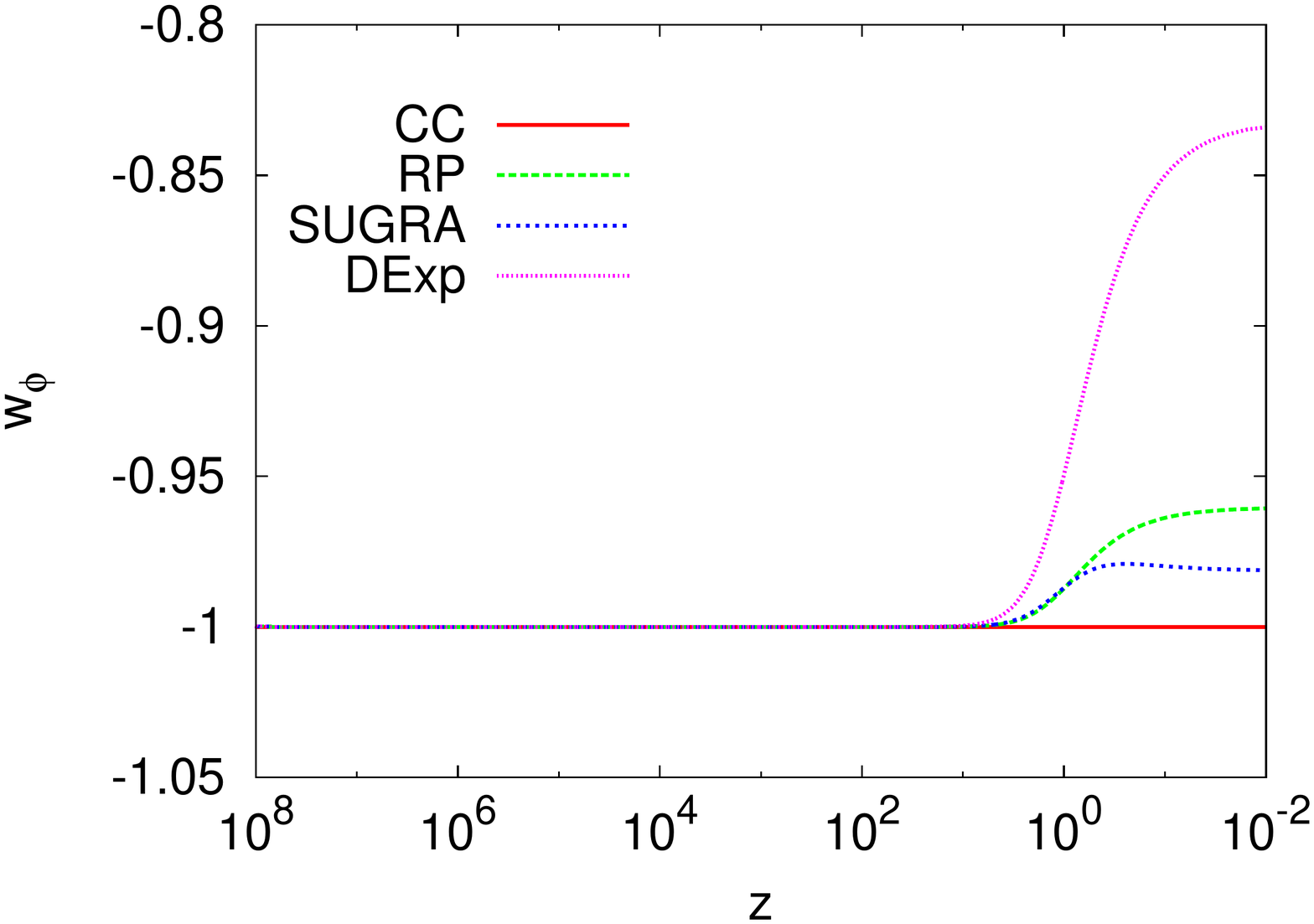}
\ff\ff \includegraphics[width=6.5cm,clip=true, trim=1cm 1cm 1cm 0cm]{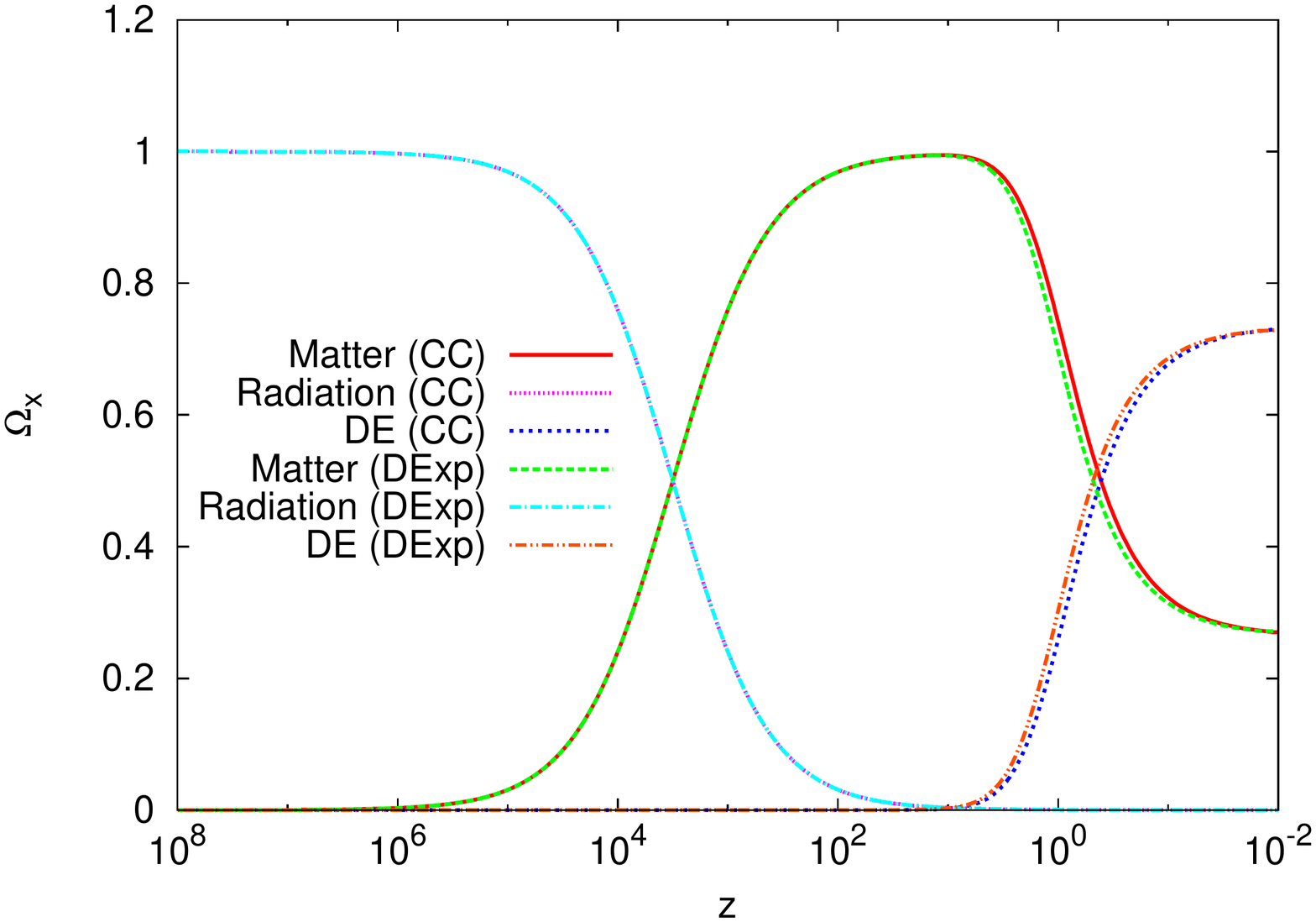}
\caption{\ii{Left:} The equation of state $w_{\phi}(z)$ of dark energy for the cosmological constant (CC), Ratra-Peebles (RP), SUGRA and double exponential (DExp) models (see section \ref{models}). \ii{Right:} The evolution of the density parameter $\Omega_x=\rho_x/\rho$ for the CC and DExp models. Apparently large deviations from $w_{\phi}=-1$ lead to subtle differences in $\Omega$ at late times.}
	\label{fig:1}
\end{figure}

We solved these coupled differential equations by integrating them numerically from redshift $z=10^{8}$ until $z=0$ (today).  Initial values were chosen by requiring that the final ($z=0$) values of the component energy densities agreed broadly with those measured today: $\Omega_{m}=0.27$, $\Omega_{r}=8.6\times10^{-5}$, $\Omega_\phi = 1-\Omega_m-\Omega_r$ and $H_{0}=70.8 \ff\mathrm{kms^{-1}Mpc^{-1}}$, where $\Omega_{x}\equiv{\rho_{x}/ \rho}.$


The evolution of $w_{\phi}$ for the four models and the background energy densities for the CC and DExp models are shown in Figure \ref{fig:1}, in which the eras of radiation, matter and dark energy domination are clearly seen in that order. The quintessence and CC models all satisfy $w_{\phi}\simeq-1$ until $z\sim1$. Despite an apparently large deviation in $w_\phi$ from $-1$ at very low redshift, the differences in the background density between the DExp and CC models are surprisingly small. Thus, it may be more fruitful to try to distinguish between quintessence and the cosmological constant via perturbations.



\section{Perturbations}

We consider only scalar perturbations to the FRW metric.  Working in the Newtonian gauge and ignoring anisotropic stresses, the perturbed metric takes the form
\begin{equation}
ds^{2}=-(1+2\Phi)dt^{2}+a(t)^{2}(1-2\Phi)dx^{i}dx_{i},
\end{equation}
where $\Phi$ is the Newtonian potential (for details see \cite{1995ApJ...455....7M}).  The components of the energy-momentum tensor, $T_\mu^{\ff\nu}$ are
\begin{center}
\begin {equation}
T_{0}^{\,\, 0}=-\rho, \,\,\,\, T_{i}^{\,\, 0}=-\frac{1}{k}(\rho+p)v_{,i}, \,\,\,\, T_{i}^{\,\, j}=p\delta_{i}^{\,\,j}, 
\end{equation}
\end{center}
where $k$ is the Fourier wavenumber, $v_{i}$ is the peculiar velocity of the perturbation for species $i$ and a comma denotes a partial derivative. 
Each energy density can be decomposed as $\rho=\overline{\rho}+\delta\rho$ where barred quantities are the background values.  We then define the variables:
\begin{center}
$\delta_{m}\equiv\dfrac{\delta \rho_{m}}{\overline{\rho}_{m}}$, \,\,\,\,   
$\delta_{r}\equiv\dfrac{\delta \rho_{r}}{\overline{\rho}_{r}}$, \,\,\,\,
$\delta\phi\equiv\phi-\overline{\phi}$.\\
\end{center}
Physically, $\delta$ is a measure of density fluctuations (overdensities and underdensities). To avoid unphysical gauge artefacts, we shall consider the gauge-invariant overdensities $\Delta_{x}$ defined as  \cite{1984PThPS..78....1K}
\setlength\arraycolsep{0.1em}
\begin{eqnarray}
\Delta_{m}&\equiv&\delta_{m}+3\left(\frac{aH}{k}\right)(1-q_{m})v_{m},\label{De1}\\
\Delta_{r}&\equiv&\delta_{r}+4\left(\frac{aH}{k}\right)(1-q_{r})v_{r},\\
\Delta_{\phi}&\equiv&\frac{\dot{\phi}\delta\dot{\phi}+(3H\dot{\phi}+{V^{\prime}}(\phi))\delta\phi-\dot{\phi}^{2}\Phi}{\rho_{\phi}},\label{De2}\\ &=&\delta_{\phi}+3\frac{aH}{k}(1+w_{\phi})v_{\phi}+\frac{3}{2}(1+w_{\phi})\left(\frac{aH}{k}\right)^{2}\overline{\Delta},\label{De3}\\
\mbox{where}\quad v_{\phi}&=&\frac{k \delta \phi}{a \dot{\phi}},\label{De4}\\
q_{x}&=&\frac{Q_{x}}{3H\rho_{x}(1+w_{x})}.
\end{eqnarray}

It will also be necessary to consider the perturbative effects arising from the dark sector interactions. We define $\epsilon$ to be the perturbation in the energy transfer $Q_x\rightarrow\widetilde{Q}_{x}$
\begin{equation}
\widetilde{Q}_{x}=Q_{x}(1+\epsilon_{x}),\label{Q1}
\end{equation}
and note that $\epsilon_x$ obeys the conservation equation
\begin{equation}
\sum_{x}{Q}_{x}\epsilon_{x}=0.\label{Q13}
\end{equation}
Similarly, the gauge-invariant variable for energy transfer, $E_x$, can be constructed as \cite{1984PThPS..78....1K}
\begin{equation}
E_{x}=\epsilon_{x}-\frac{a\dot{Q}_{x}}{kQ_{x}}v_{x}.\label{Q2}
\end{equation}


\subsection{Covariant formulation of dark sector interactions}

We now describe how dark sector interactions can be covariantly formulated and incorporated into the gauge-invariant system (see \cite{2008JCAP...07..020V, 2010MNRAS.402.2355V, PhysRevD.80.063530} for previous attempts in this direction for simple dark energy models). 


Dark sector interactions can be covariantly described by the conservation equation 
\beq
 \nabla_\nu T^{\mu\nu}_x = Q_x^{\mu}.
 \eeq
The 4-vector $Q^{\mu}_x$ can be constructed using the function $Q_x$ by
\beq
{Q}_{(x)\mu}={Q}_{(x)}{u}_{(a)\mu},\label{myq} 
\eeq
where $u_{(a)\mu}$ is the 4-velocity of component $a$ (which may be different from $x$).  By perturbing \re{myq}, we find
\begin{equation}
\widetilde{Q}_{(x)\mu}=\widetilde{Q}_{(x)}\widetilde{u}_{\mu}+\widetilde{f}_{(x)\mu}, \label{Q3}
\end{equation} 
where $u_{\mu}$ is the average velocity.  The perturbed average 4-velocity ${u}_\mu\rightarrow\widetilde{u}_\mu$ and perturbed 4-velocity of component $a$ (${u}_{(a)\mu} \rightarrow \widetilde{u}_{(a)\mu}$) can be expressed as
\begin{equation}
\widetilde{u}_{0}=\widetilde{u}_{(a)0}=-a(1+\Phi), \qquad \widetilde{u}_{j}=a\bar{v}{Y_j}, \qquad  \widetilde{u}_{(a)j}=av_a{Y_j}.\label{pert4v}
\end{equation}
where $Y_{j}$ is a basis vector constructed from harmonic functions\footnote{see Appendix C of \cite{1984PThPS..78....1K} for detail. We will not require the explicit form of $Y_j$ in the evolution equations.}.  In \ref{Q3}, the vector $\widetilde{f}_{(x)\mu}$ can be interpreted as the momentum exchange with components given by 
\begin{equation}
\widetilde{f}_{(x)0}=0, \qquad \widetilde{f}_{(x)j}=aH\rho_{x}(1+w_{x})f_{x}Y_{j},\label{Q14}
\end{equation}
where $f_x$ is an ordinary scalar representing the amplitude of the momentum exchange. In addition, $f_x$ satisfies the conservation of momentum
\begin{equation}
\sum_{x}\rho_{x}(1+w_{x})f_{x}=0.\label{Q8}
\end{equation}
As with the energy exchange, we can similarly recast $f_x$ in a gauge-invariant form $F_x$
\begin{equation}
F_{x}\equiv f_{x}-\frac{Q_{x}(v_{x}-\bar{v})}{H\rho_{x}(1+w_{x})}.\label{Q7}
\end{equation}

All the ingredients introduced so far are now sufficient to allow us to study the evolution of the perturbation variables $\{\Delta_m,\Delta_r,\Delta_\phi,v_m,v_r,v_\phi\}$. In the regime where perturbations are small, we find the following set of coupled differential equations  \cite{1984PThPS..78....1K}
\begin{eqnarray}
\frac{d\Delta_{m}}{dN}&=&\frac{9}{2}\frac{aH}{k}(1+w)(\bar{v}-v_{m})-\frac{Q_{m}\Delta_{m}}{aH\rho_{m}}-\frac{k}{aH}v_{m}+\frac{Q_{m}E_{m}}{\rho_{m}H}+F_{m},\label{Dm}\\
\frac{d\Delta_{r}}{dN}&=&6\frac{aH}{k}(1+w)(\bar{v}-v_{r})+\Delta_{r}-\frac{4k}{3aH}v_{m},\nonumber\\
\\
\dfrac{d\Delta_{\phi}}{dN}&=&\left(3w_{\phi}-\frac{Q_{\phi}}{H\rho_{\phi}}\right)\Delta_{\phi}+\nonumber\\
&& \frac{2x^{2}}{\rho_{\phi}}\left[\frac{9}{2}\frac{aH}{k}(1+w)(\bar{v}-v_{\phi})-\frac{k}{aH}v_{\phi}\right]+\frac{Q_{\phi}E_{\phi}}{\rho_{\phi}H}+(1+w_{\phi})F_{\phi},\nonumber\\
\label{Dp}\\
\frac{dv_{m}}{dN}&=&-v_{m}-\frac{3aH}{2k}\bar{\Delta}+F_{m},\label{vem}\\
\frac{dv_{r}}{dN}&=&-v_{r}+\frac{k}{4aH}\Delta_{r}-\frac{3aH}{2k}\bar{\Delta},\\
\frac{dv_{\phi}}{dN}&=&-v_{\phi}+\frac{k}{(1+w_{\phi})aH}\Delta_{\phi}-\frac{3aH}{2k}\bar{\Delta}+F_{\phi},\label{vep}
\end{eqnarray}
where
\begin{eqnarray}
\bar{v}&=&\frac{1}{1+w}\left(2x^{2}v_{\phi}+\frac{4}{3}r^{2}v_{r}+m^{2}v_{m}\right),\\
\bar{\Delta}&=&\left(m^{2}\Delta_{m}+r^{2}\Delta_{r}+(x^{2}+y^{2})\Delta_{\phi}\right)+\frac{a}{k}\sum_{x}Q_{x}v_{x},\\
w&=&\frac{1}{\rho}(\rho_{r}w_{r}+\rho_{\phi}w_{\phi})=\frac{r^{2}}{3}+(x^{2}+y^{2})w_{\phi}.\label{w3}
\end{eqnarray}


Note that by setting $Q_{x},E_{x}$ and $F_{x}$ to 0, we recover the non-interacting case. The initial values of the perturbations are fixed using adiabatic initial conditions \cite{2010MNRAS.407.1989C, 2003PhRvD..68f3505D} which we calculate to be
\begin{center}
$\dfrac{\Delta_{m}}{1-q_{m}}=\dfrac{3}{4}\Delta_{r}$, \,\,\,\,
$\Delta_{\phi}=3\dfrac{\dot{\phi^{2}}}{\rho_{\phi}}\left(\dfrac{aH}{k}\right)^{2}\overline{\Delta}$, \,\,\,\,
$q_{x}=\dfrac{Q_{x}}{3aH\rho_{x}}$
\begin{equation}
v_{m}=\dfrac{k}{3aH}\dfrac{\Delta_{m}}{1-q_{m}}, \,\,\,\,\,
v_{r}=\dfrac{k\Delta_{r}}{3aH}, \,\,\,\,\,
v_{\phi}=0.\label{Ad}
\end{equation}
\end{center}

The only free variable is the initial matter density perturbation which is determined by normalising the matter power spectrum at a pivot scale which we take to be $k=0.001 \mbox{Mpc}^{-1}$. The power spectrum of energy component $x$ is calculated by comparing the amplitudes of the growing mode at early and late times 
\begin{equation}\label{Ps}
P_{x}(k)=k^{n_{s}}\left[\frac{\Delta_{x}(z=0)}{\Delta_{x}(z=10^{8})}\right]^{2},
\end{equation}
where $n_s$ is the primordial scalar spectral index taken to be $0.96$.

\subsection{Power spectra without interactions}

\begin{figure}[ht!]
	\centering
		\includegraphics[width=7 cm]{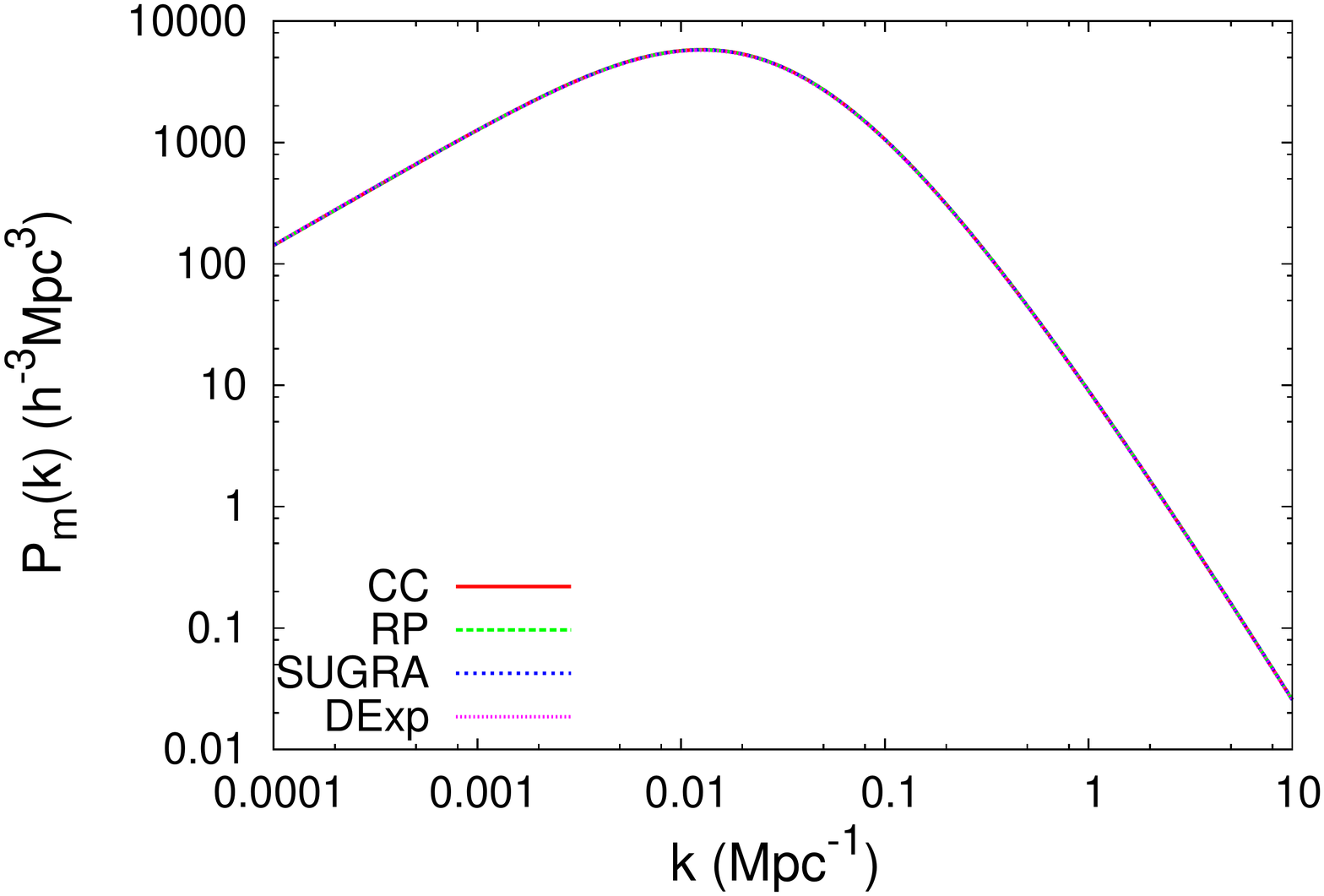}
	 		\ff\ff \includegraphics[width=7 cm]{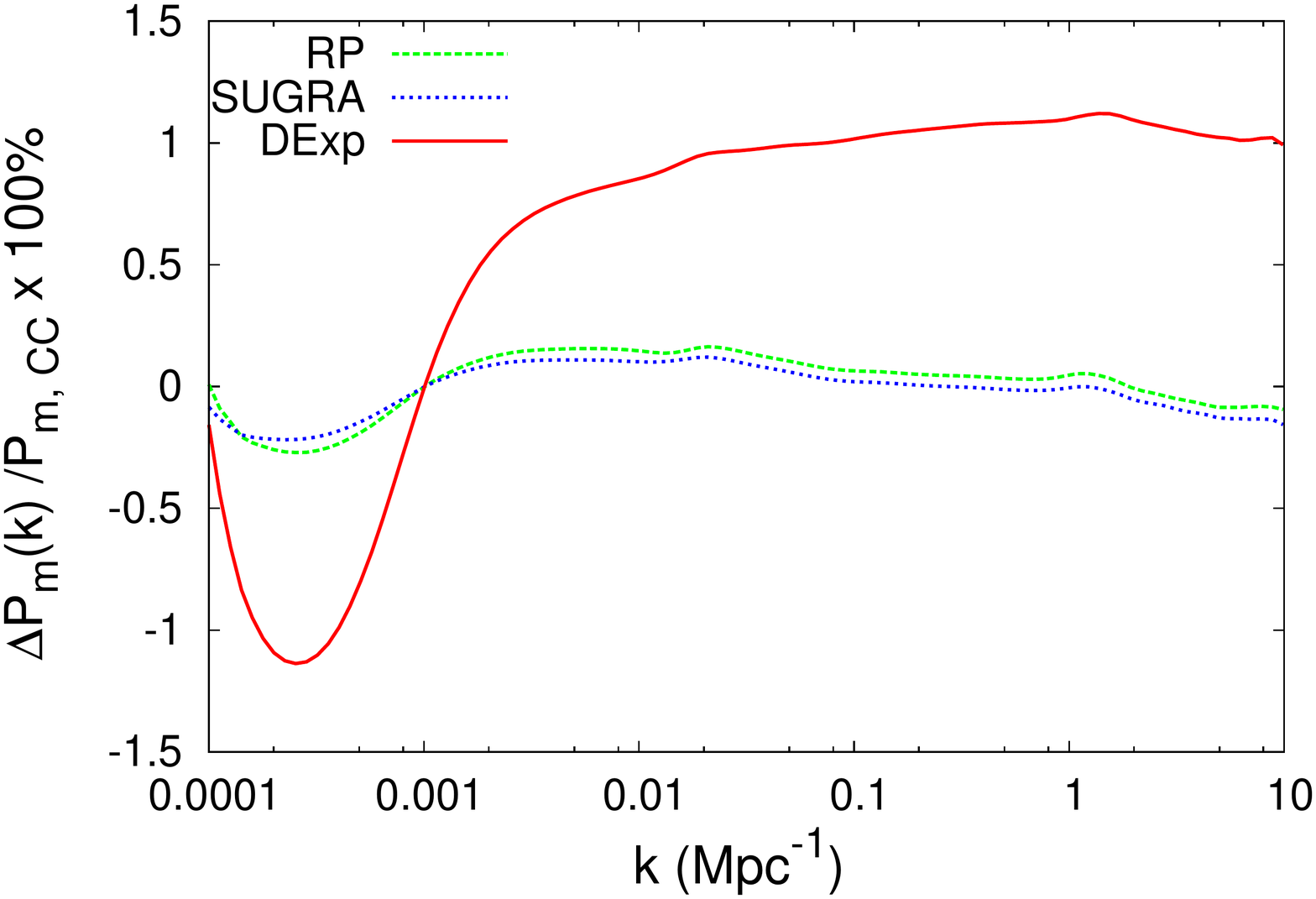}
	\caption{\ii{Left:} If there is no interaction in the dark sector, the linear matter power spectra for all four `best-fit' dark energy models are essentially indistinguishable. \ii{Right:} the fractional percentage difference between the matter power spectra of the quintessence models and the cosmological constant. The best-fit models all exhibit sub-percent differences from the cosmological constant.}
	\label{fig:2}
\end{figure}

Figure$\nolinebreak \,\ref{fig:2}$ shows the matter power spectrum for all four dark energy models as well as the fractional difference in the matter power spectra for the quintessence models relative to the \CC expressed as a percentage.  
$$ \frac{P_{m}(k)^{\mathrm{model}}-P_{m}(k)^{CC}}{P_{m}(k)^{CC}} \times 100\%.$$
Note that our calculations are not valid for scales $k>0.1$ where we expect a significant enhancement in the power spectra due to non-linearities. The equations governing the perturbations in the non-linear regime are beyond the scope of our work. To study the perturbations in this regime, one requires an $N-$body simulation as in \cite{2010MNRAS.401..775A, Rasera:2010ar} or a more sophisticated perturbation theory \cite{Bernardeau:2001qr}.

\begin{figure}[t!]
	\centering
		\includegraphics[width=9cm]{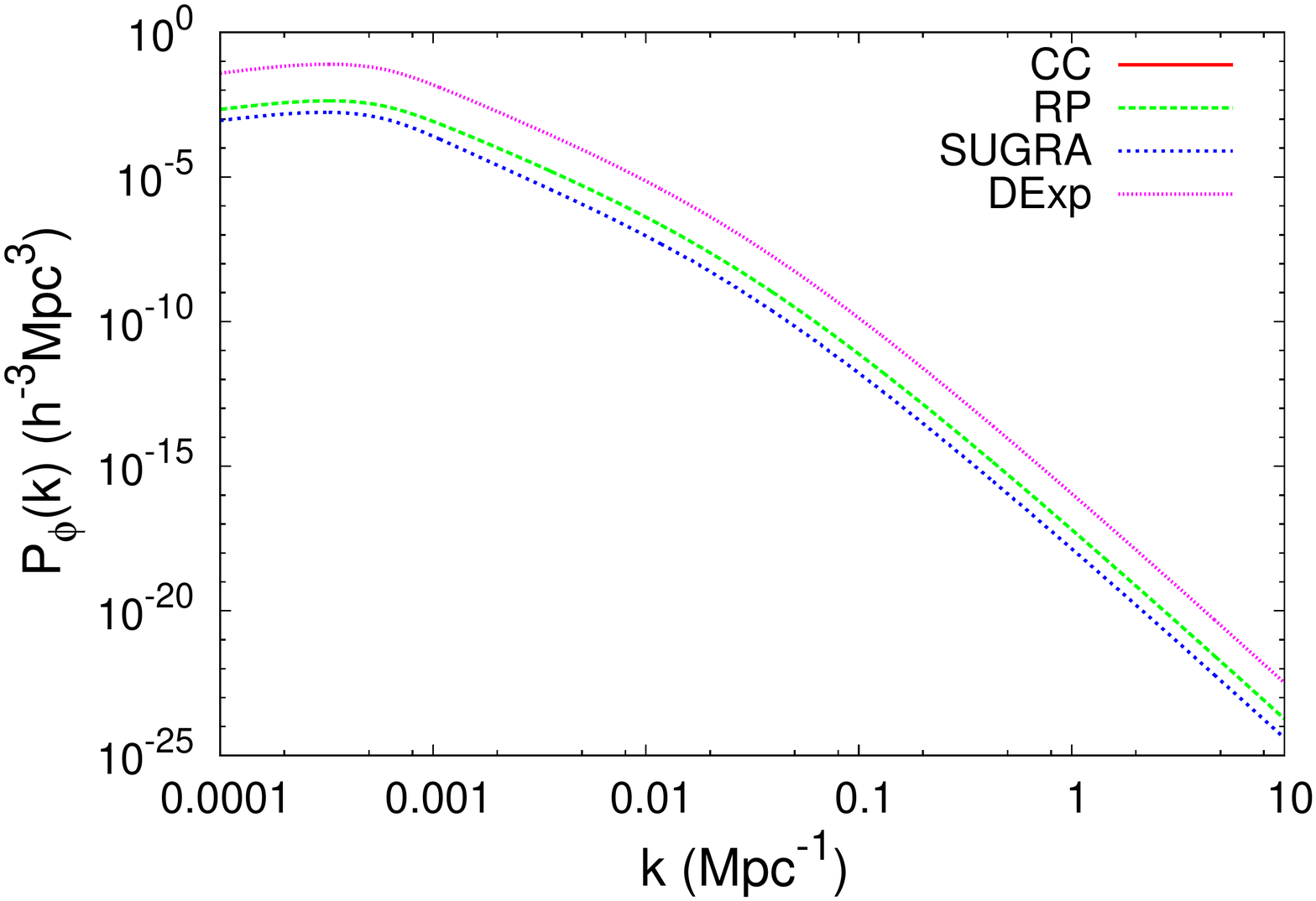}
	\caption{The linear power spectra for dark energy for the three quintessence models with no interactions. Note the `turnover' on very large scales close to the Hubble radius ($\sim 2\times10^{-4} \mathrm{Mpc}^{-1}$) indicating the typical size of dark energy perturbations.}
	\label{fig:5}
\end{figure}

Differences between the models are small with the largest difference $\approx 1\%$ observed for the DExp potential.  The quintessence models have an excess power for $k > 0.001 \mathrm{Mpc^{-1}}$ and a power deficit for $k < 0.001 \mathrm{Mpc^{-1}}$.  This correlates well with the `turnover'  to the dark energy spectra shown in Figure$\nolinebreak \,\ref{fig:5}$. 
Nevertheless, these small differences of order $1\%$ in matter power spectra for the models are essentially unobservable.

The power spectrum for dark energy perturbations, shown in Figure$\nolinebreak \,\ref{fig:5}$, are roughly four orders of magnitude smaller than the matter power spectra. Thus, we do not expect to see large differences in the matter power spectra of the quintessence models due to dark energy clustering.  We see that the large-scale clustering of dark energy enhances the clustering of dark matter on large scales.  The dark energy power spectra are approximately constant on very large scales ($k<0.001 \mathrm{Mpc^{-1}}$), and decay exponentially on small scales. In fact, dark energy clustering is significant mainly on large scales in all models in which the sound speed of dark energy equals the speed of light \cite{2010PhRvD..81j3513D}. The decay of the power spectrum is split into two regions with different exponents changing near the turnover scale in the matter power spectrum.  The differences in the power spectra of the different quintessence models is due to the  different values of $w_{\phi}$ at late times, since $\Delta_{m}\propto (1+w)$ (see $\nolinebreak \,\ref{Dm}$ and $\ref{w3}$). We will shortly investigate the effects of dark sector interactions on these power spectra.





\section{Integrated Sachs-Wolfe effect}

The Integrated Sachs-Wolfe effect (ISW) is an effect observed in the CMB for small multipole number $\ell$ caused by the blueshifting of photons as  the gravitational potential wells become shallower due to the late-time cosmic acceleration \cite{1967ApJ...147...73S}. The gravitational potential $\Phi$ can be decomposed as:
\begin{equation}
\Phi=\Phi_{m}+\Phi_{r}+\Phi_{\phi}.
\end{equation}
The potential $\Phi$ is related to the perturbation variables by the cosmological Poisson's equation
\begin{equation}
\Phi=-\frac{3}{2}\left(\frac{aH}{k}\right)^{2}\overline{\Delta}.
\end{equation}


Consider the CMB temperature anisotropy ${\Delta T/ T }(\mb{n})$ along the line-of-sight unit vector $\mb{n}$. The CMB anisotropy power spectrum is given by
\begin{equation}
\left\langle \frac{\Delta T}{T}(\textbf{n})\frac{\Delta T}{T}(\textbf{m}) \right\rangle_{\bf {n}\cdot \bf {m}=\mu}=\frac{1}{4\pi}\sum_{\ell}(2\ell+1)C_{\ell}P_{\ell}(\mu),
\end{equation}
where $C_\ell$ is amplitude of the CMB angular power spectrum at the $\ell$-th multipole and $P_\ell$ is the Legendre polynomial of order $\ell$. By decomposing $\Delta T/T$ into the spherical harmonics and integrating over all directions we find \cite{2008cmb..book.....D} 
\begin{eqnarray}\label{Cl2}
{C_{\ell}^{\mathrm{ISW}}\sim \int_{0}^{\infty} \int_{z_{\mbox{\scriptsize dec}}}^{0} k^{2}} \left| -\frac{H}{a}\frac{d\Phi}{dz} J^{2}_{\ell}\left(k\int_{0}^{z}\left[\frac{1}{H(z^\pr)(1+z^\pr)}\right]dz^\pr\right)\right|^{2}dz\, dk,
\end{eqnarray}
where $z\sub{dec}$ is the redshift at which photon decoupling occurs ($\sim1020$) and  $J_{\ell}(x)$ is the spherical Bessel function of order $\ell$.

\begin{figure}[t!]
	\centering
		\includegraphics[width=9cm]{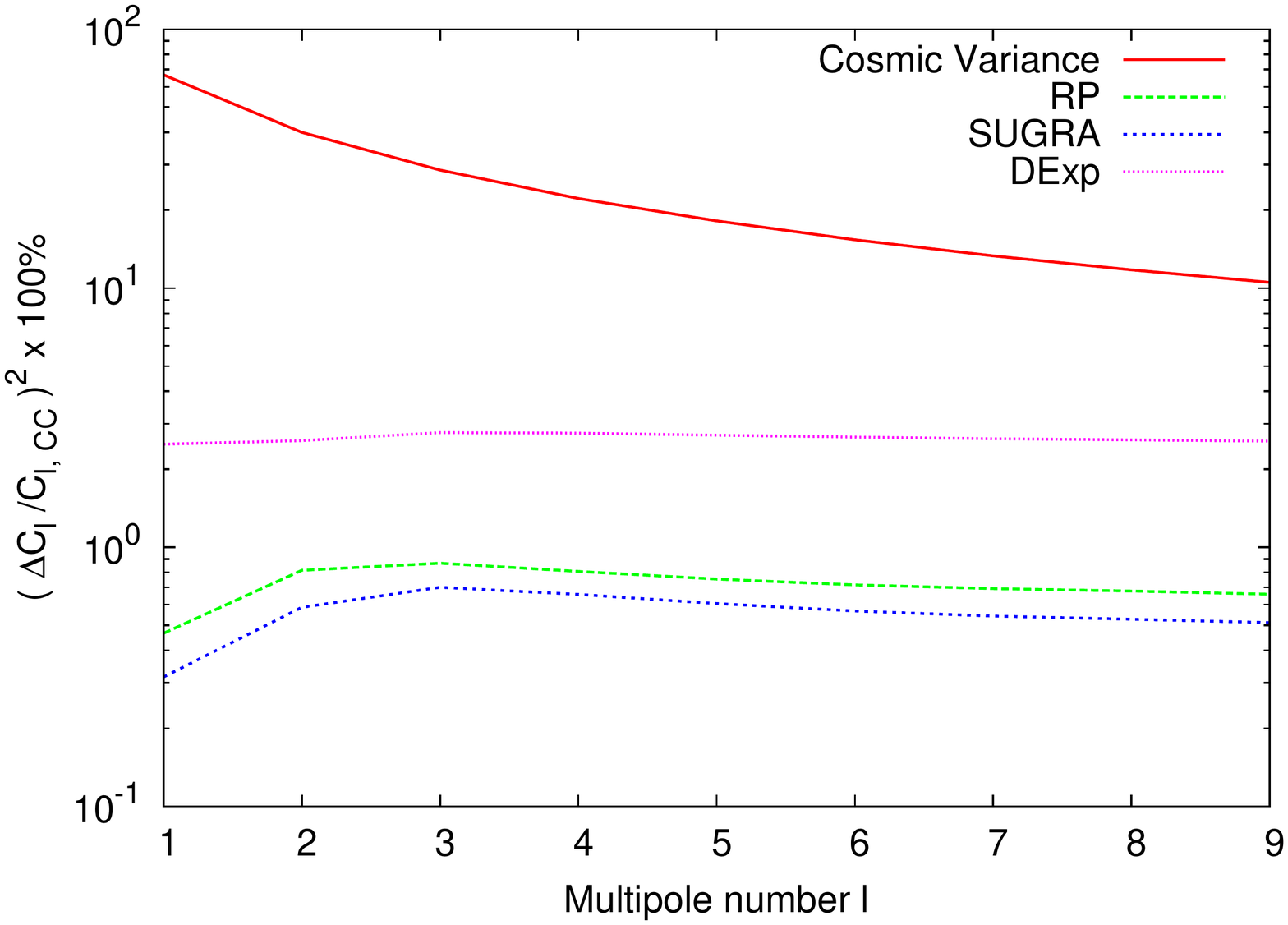}
	\caption{The percentage differences between $C_{\ell}^{\mathrm{ISW}}$ for the three quintessence models and the CC model. The contributions from quintessence are overwhelmed by cosmic variance, hence the models are observationally indistinguishable via the ISW effect.}
	\label{fig:6}
\end{figure}

Figure$\nolinebreak \,\ref{fig:6}$ shows the squared percentage differences between $C_{\ell}^{\mathrm{ISW}}$ for the three quintessence models and the \CC model
$$\left(\frac{C_{\ell}^{\mathrm{quint}}-{C_{\ell}^{\mathrm{CC}}}}{C_{\ell}^{\mathrm{CC}}}\right)^{2}\times 100\%,$$ in comparison with the cosmic variance
\begin{equation}
(C_{\ell}^{\mathrm{var}})^{2}=\frac{2}{2\ell+1}.
\end{equation}  

The relative differences in the ISW effect for quintessence models and $\Lambda$CDM are completely overwhelmed by cosmic variance. They are also much smaller than indicated in other investigations in which the equation of state for dark energy were parameterised by simple ansatz e.g. $w_{\phi}=w_0$ or $w_{\phi}=w_0+ w_1z$ (where $w_0$ and $w_1$ are constants). This is perhaps not surprising since our `best-fit' quintessence models can closely replicate $\Lambda$CDM dynamics whereas these parametrizations can diverge from $\Lambda$CDM dynamics substantially.

In summary, we find that the ISW is an ineffective discriminant between the cosmological constant and models of quintessence that deviate from $w=-1$ at very low redshifts. In the next section, we investigate whether this conclusion is changed if there are interactions between quintessence and dark matter.

\section{Modelling dark sector interactions}
In this section, we introduce three models of dark sector interactions and explain how they can be rewritten in covariant forms.


\subsection{Dark matter decay}

We first consider a dark sector interaction of the form
\begin{equation}
Q_{m}=-A\rho_{m},
\end{equation}
which represents the decay of dark matter into dark energy with a constant decay rate.  The decay rate per unit energy density of dark matter is quantified by the positive constant $A$, which we refer to as the interaction strength.  This form of interaction has been previously investigated in \cite{2008JCAP...07..020V}.  

In the covariant formulation, the interaction term is given by $A\rho_{m}u_{m}$, where $u_{m}=(-a,\bf{0})$, since this is the rate at which 4-momentum is transferred to dark energy.  Hence,
\begin{equation}
Q_{(m)\mu}=-A\rho_{m}u_{(m)\mu}.
\end{equation}
We perturb the above equation to find
\begin{equation}
\widetilde{Q}_{(m)\mu}=Q_{m}(1+\delta_{m})\widetilde{u}_{(m)\mu}, \qquad \widetilde{Q}_{\phi}=Q_{\phi}(1+\delta_{m})\widetilde{u}_{(m)\mu}.
\end{equation}
Using Equations$\nolinebreak \, \ref{De1}$, $\ref{Q1}$, $\ref{Q2}$, \ref{pert4v} and the above, we find the gauge-invariant energy transfer
\begin{equation}
\epsilon_{m}=\delta_{m}, \qquad E_{m}=\Delta_{m}, \qquad E_{\phi}=\Delta_{m}+3\frac{aH}{k}\bkt{1-\frac{Q_{m}}{3H\rho_{m}}}(v_{\phi}-v_{m}).
\end{equation}
Similarly, using Equations$\nolinebreak \,\ref{Q3}$, $\ref{Q14}$, $\nolinebreak \,\ref{Q8}$ and $\ref{Q7}$, we find the momentum transfer
\begin{equation}
f_{m}=\frac{Q_{m}(v_{m}-\bar{v})}{H\rho_{m}}, \qquad F_{m}=0, \qquad F_{\phi}=\frac{Q_{\phi}(v_{m}-v_{\phi}).}{H\rho_{\phi}(1+w_{\phi})}.
\end{equation}


\subsection{Dark energy decay}

The second model represents the case in which dark energy decays into dark matter at a constant decay rate \cite{1996PhRvD..54.2571L}. This may be applicable, for instance, in the early dark energy scenario \cite{2006JCAP...06..026D}.  The interaction term has the form
\begin{equation}
Q_{\phi}=-A\rho_{\phi}, \qquad Q_{m}=A\rho_{\phi}, \qquad Q_{(\phi)\mu}=-A\rho_{\phi}u_{(\phi)\mu}.\label{Q20}
\end{equation}  
Similarly, using Equations$\nolinebreak \,\ref{Q1}$--$\ref{Q7}$, we find the gauge-invariant energy and momentum transfers 
\begin{equation}
E_{\phi}=\Delta_{\phi}-3\frac{aH}{k}(1+w_{\phi})\left[\frac{Q_{\phi}v_{\phi}}{3H\rho_{\phi}(1+w_{\phi})}+\frac{aH}{2k}\overline{\Delta}\right],
\end{equation}
\begin{equation}
E_{m}=\Delta_{\phi}+3\frac{aH}{k}(1+w_{\phi})\left[v_{m}-v_{\phi}-\frac{Q_{\phi}v_{m}}{3H\rho_{\phi}(1+w_{\phi})}-\frac{aH}{2k}\overline{\Delta}\right].
\end{equation}
\begin{equation}
F_{\phi}=0, \qquad F_{m}=\frac{Q_{m}(v_{\phi}-v_{m})}{H\rho_{\phi}}.
\end{equation}
Note that the momentum transfer $F_m$ and $F_\phi$ are `orthogonal' to those in the dark matter decay model.

\subsection{Scalar-tensor model}

Finally, let us consider a interaction term which appears in some scalar-tensor theories of gravity \cite{2004PhRvD..69j3524A} 
\begin{equation}
Q_{(\phi) \mu}=A\rho_{m}\nabla_{\mu}\phi, \qquad Q_{(m)\mu}=-A\rho_{m}\nabla_{\mu}\phi.\label{Q15}
\end{equation}  
Using Equation $\ref{Q3}$ we find 
\begin{equation}
\widetilde{Q}_{\phi}=-\widetilde{u}^{\mu}A \rho_{m}(1+\delta_{m}) \partial_{\mu}\widetilde{\phi}, \qquad Q_{\phi}=A\rho_{m}\dot{\phi} \label{Q10}
\end{equation}
where $u^{\mu}$ is the average velocity.  To calculate the gauge-invariant interaction, we use the following useful expressions for the perturbed scalar field
\begin{equation}
\frac{\partial \widetilde{\phi}}{\partial t}=\dot{\phi}-\Phi\dot{\phi}+\delta\dot{\phi}, \quad \quad \delta\phi=\frac{av_{\phi}\dot{\phi}}{k},
\end{equation} 
\begin{equation}
\frac{\delta\dot{\phi}}{\dot{\phi}}=\frac{\Delta_{\phi}}{1+w_{\phi}}+\Phi-\frac{(3H\dot{\phi}+V')\delta\phi}{\dot{\phi}^{2}}.
\end{equation}

Using Equations$\nolinebreak \, \ref{Q5}$, $\ref{pert4v}$ and $\nolinebreak \ref{Q13}$, we find the gauge-invariant energy transfer variables
\begin{equation}
E_{\phi}=\Delta_{m}+\frac{\Delta_{\phi}}{1+w_{\phi}}+\frac{3aH}{k}\left(\frac{Q_{m}}{3H\rho_{m}}-1\right)v_{m}+\frac{3}{2}\left(\frac{aH}{k}\right)^{2}\overline{\Delta},
\end{equation}
\begin{equation}
E_{m}=\Delta_{m}+\frac{\Delta_{\phi}}{1+w_{\phi}}+\frac{3aH}{k}\left(\frac{Q_{m}v_{m}}{3H\rho_{m}}-v_{\phi}\right)+\frac{3}{2}\left(\frac{aH}{k}\right)^{2}\overline{\Delta}-\frac{3aH}{k}\frac{y^{2}\lambda}{\sqrt{6}x}(v_{m}-v_{\phi}).
\end{equation}

Similarly, from $\nolinebreak \,\ref{De4}$, $\ref{Q14}$, $\nolinebreak \ref{Q7}$ and $\ref{Q15}$, we find the gauge-invariant momentum transfer variables to be
\begin{equation}
f_{\phi}=\frac{A\rho_{m}\dot{\phi}(v_{\phi}-\bar{v})}{H\rho_{\phi}(1+w_{\phi})}, \qquad F_{\phi}=0, \qquad F_{m}=\frac{Q_{\phi}(v_{m}-v_{\phi})}{H\rho_{m}}.
\end{equation}

\section{Results}

We are now ready to calculate the observables for the different interaction models. In particular, we would like to establish if dark sector interactions could give rise to observable imprints in the power spectrum or the ISW effect whilst satisfying the constraint that $w$ is very close to $-1$. We use the SUGRA potential for the dark energy decay and scalar-tensor model because, at the background level, it behaves most similarly to the cosmological constant (see Figures$\nolinebreak \,\ref{fig:1}$ and $\ref{fig:2}$).  It is computationally more challenging to implement the SUGRA potential with the dark matter decay model due to the non-monotonic nature of the potential and the RP potential was used instead (where an analytic solution for $\phi$ can be calculated given the value of $V(\phi)$). In all cases, however, we did not find a sensitive dependence on the choice of potential.

\subsection{Instability in the dark energy decay model}

\begin{figure}[ht]
	\centering
		\includegraphics[width=12cm]{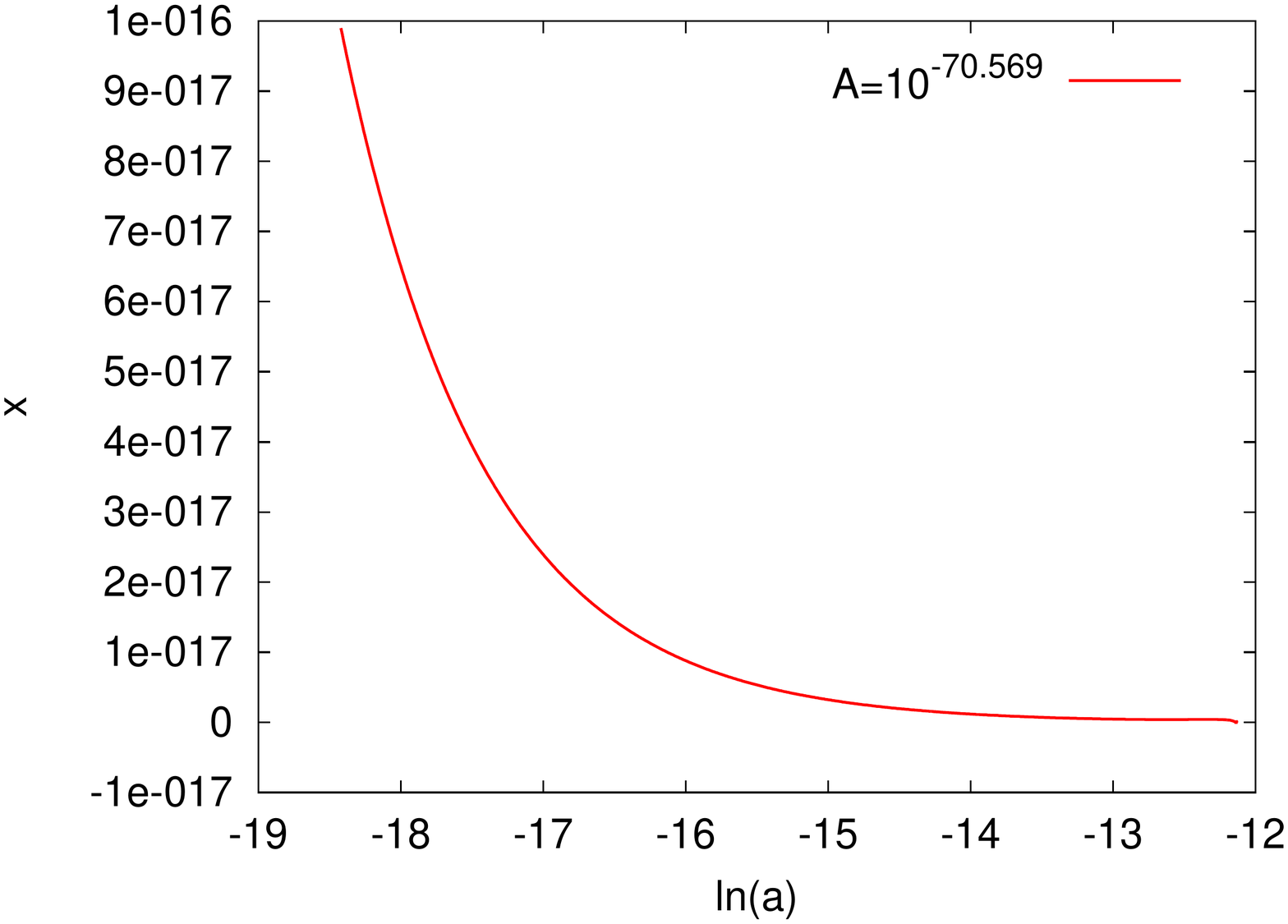}
	\caption{Evolution of the variable $x\propto\dot\phi$ in the decaying dark energy model using the SUGRA potential.  $x$ passes through zero and becomes negative at $\ln a=-12.336$ causing an instability in the power spectra. This behaviour occurs for interaction strengths larger than $\sim10^{-70}\mpl$.}
	\label{fig:7}
\end{figure}

Our computation shows that the decaying dark energy model was found to suffer from an instability in the perturbations.  From Equation$\nolinebreak \,\ref{Q16}$, we see that significant values of interaction strength act to drive the value of $x$ towards 0, causing ${dx}/{dN}$ and $v_{\phi}$ to become infinite.  Below the critical value of interaction strength associated with this instability, there is no significant effect on the power spectra or ISW effect.  An example of this instability is shown in Figure$\nolinebreak \,\ref{fig:7}$  in which the interaction strength $A$ is chosen to be just above the critical value.  Starting at $z=10^{8}$, we see that $x$ decreases to 0 at $\ln a=-12.3$ where the instability develops.  By considering very early times, we can effectively set the limit on the interaction strength to 0 and rule out this model of dark energy.  This is because even small interaction strengths would, at early times,  cause the small amount of scalar field kinetic energy to quickly dissipate, leading to this instability.  In general, regardless of the choice of quintessence potential, we find that the susceptibility of the dark energy decay model to this instability makes it unviable as a model for interacting dark energy.


%

\subsection{Dark matter decay}

\begin{figure}[ht]
	\centering
		\includegraphics[width=12cm]{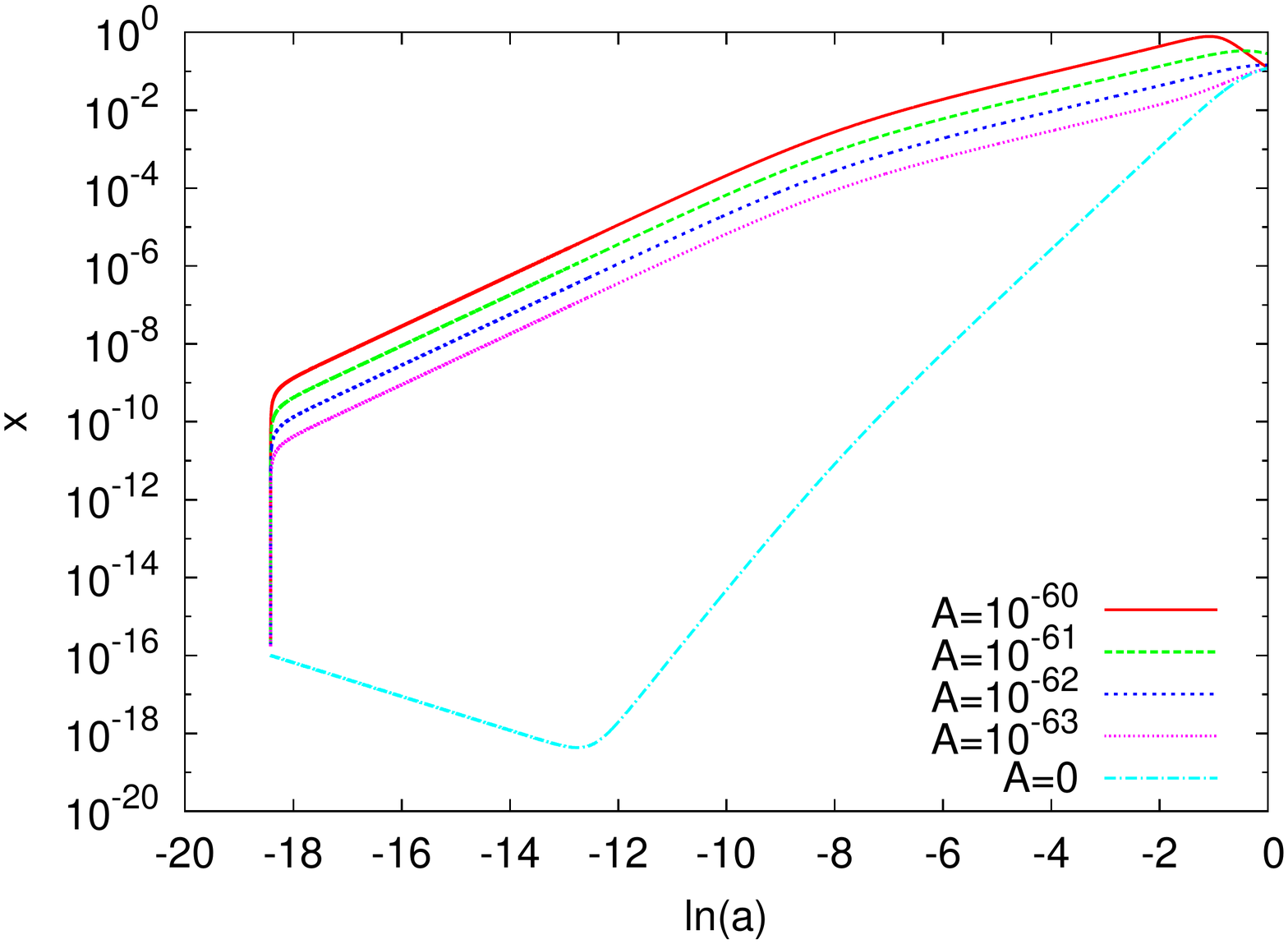}
	\caption{Evolution of the energy variable $x \propto \dot \phi$ in the decaying dark matter model using the RP potential.  We see that increasing the interaction strength $A$ boosts the value of $x$ at early times which can lead to an increase in late time dark energy densities.}
	\label{fig:8}
\end{figure}

The primary effect of the dark matter decay model is to alter the time at which dark energy comes to dominate the Universe.  This is because dark matter decay is most effective at early times when matter dominates over dark energy.  This means that the background evolution of our `best-fit' models is sensitive to the interaction strength.  For instance, an interaction strength $A=10^{-62}\mpl$ changes the dark energy density at $z=0$ by roughly $1\%$, but leaves the matter power spectrum and ISW effect virtually unchanged.  



\begin{figure}[ht]
	\centering
		\includegraphics[width=12cm]{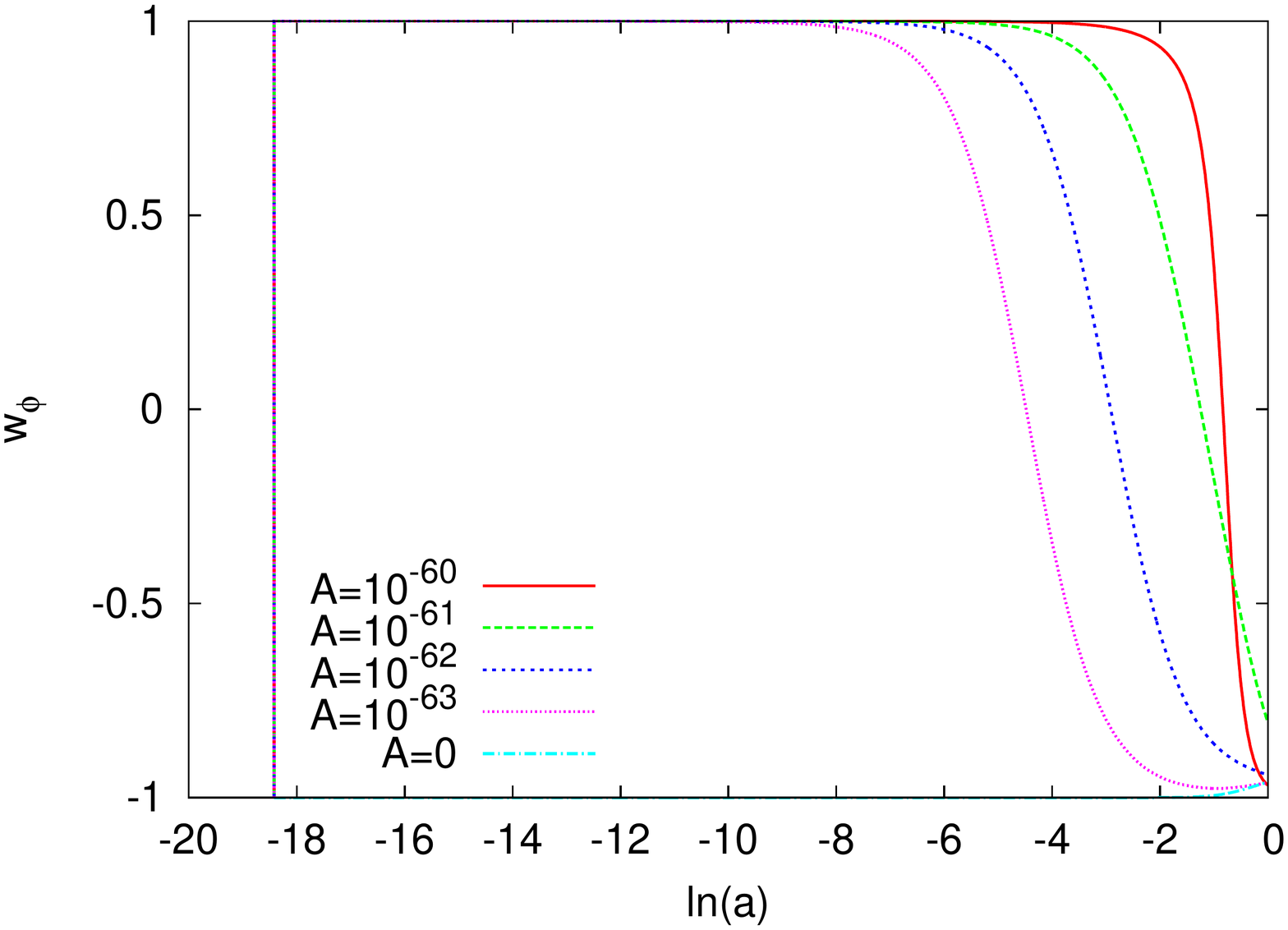}
	\caption{Evolution of $w_{\phi}$ for the dark matter decay models presented in Figure $\ref{fig:8}$.  At early times  $w_{\phi}$ responds to the sharp increase in $x$ by rising suddenly to  a value of 1. See the text for further discussion.}
	\label{fig:9}
\end{figure}

Figure $\ref{fig:8}$ shows the evolution of the $x$ variable for interaction strengths in the range $10^{-63}-10^{-60}\mpl$.  We see that the dark energy density is sharply boosted at early times, as the energy is immediately transferred from dark matter  to dark energy.  In Figure $\ref{fig:9}$, we show the evolution of $w_{\phi}$ over the same range of interaction strengths.  We see the corresponding sharp rise in $w_{\phi}$ towards 1 at early times.  This has the effect of redshifting away the excess dark energy at a rate $\rho_{\phi}\propto a^{-6}$.  This rapid redshifting means that for a wide range of $V(\phi)$, dark energy does not come to dominate the universe too soon and so this model of interactions can indeed be reconciled with observations.  

A number of previous investigations of this interaction have used simple parameterisations of $w_{\phi}$.  However, we see from Figure $\ref{fig:8}$ that the complex redshifting behaviour of dark energy would be extremely difficult to reproduce with a simple analytic form of $w_{\phi}$.  In fact, an instability in this form of interaction was found in \cite{2008JCAP...07..020V} when considering dark energy parametrized by $w_{\phi}=\mathrm{constant}$.  We can see that in a $w=$constant model where $w<0$, the early boost in the dark energy density will not be redshifted away relative to the dark matter energy density and dark energy will come to dominate the Universe too soon for any appreciable interaction strength.  It is interesting that due to the redshifting behaviour described above, our quintessence models do not encounter such an instability.


\subsection{Scalar-tensor model}

\begin{figure}[ht!]
	\centering
		\includegraphics[width=15cm]{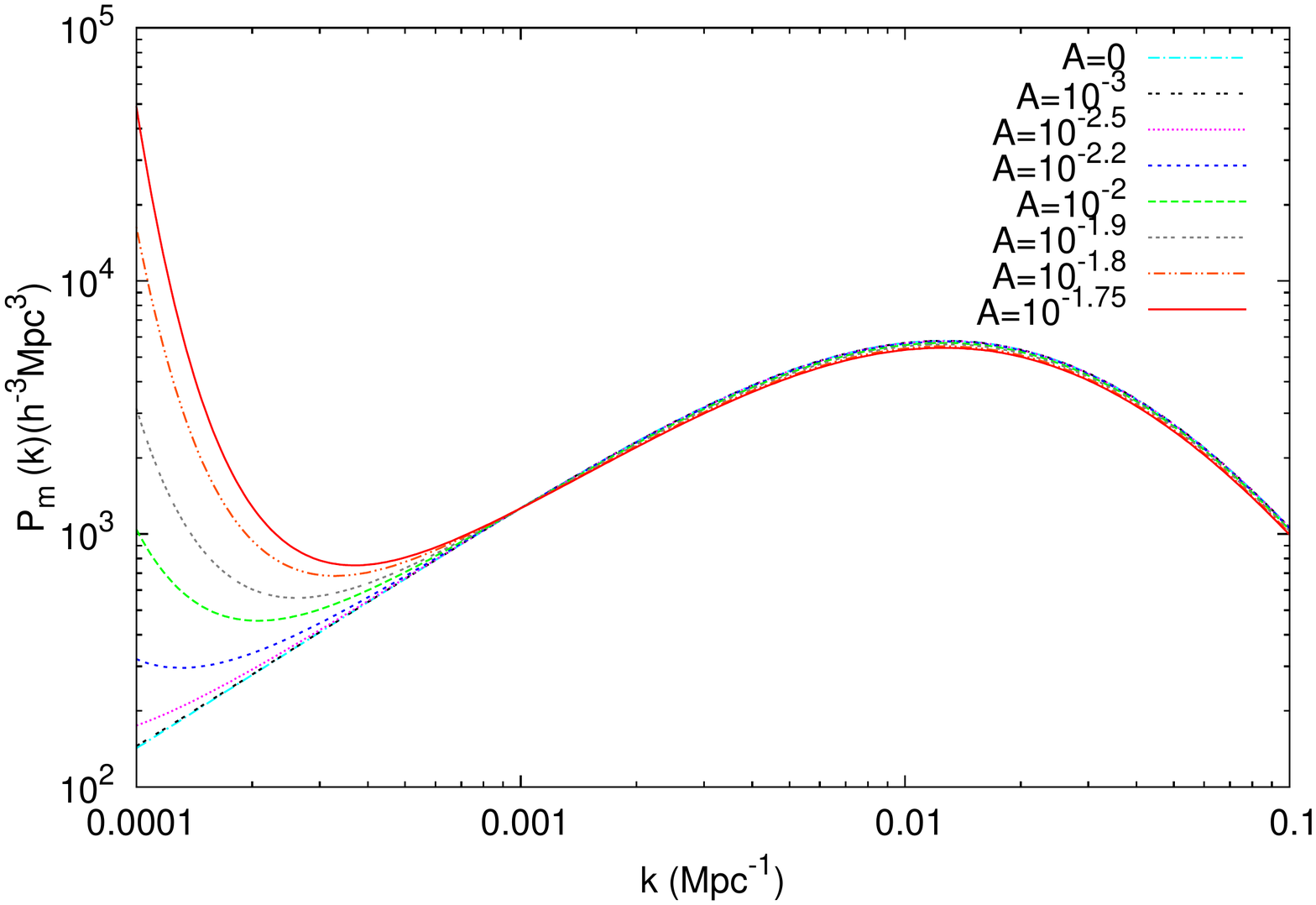}
	\caption{The linear matter power spectra in the SUGRA quintessence model with dark sector interaction of the `scalar-tensor' form $Q_{\phi}=A\rho_m\dot{\phi}$.  Large values of $A$ cause a significant enhancement to the power spectrum on very large scales. The interaction also leads to a smaller $\sim 6\%$ difference between the power spectra of the largest interaction strength and no interaction on scales $k>0.01 \mathrm{Mpc^{-1}}$.}

	\label{fig:10}
\end{figure}

\begin{figure}[ht!]
	\centering
		\includegraphics[width=15cm]{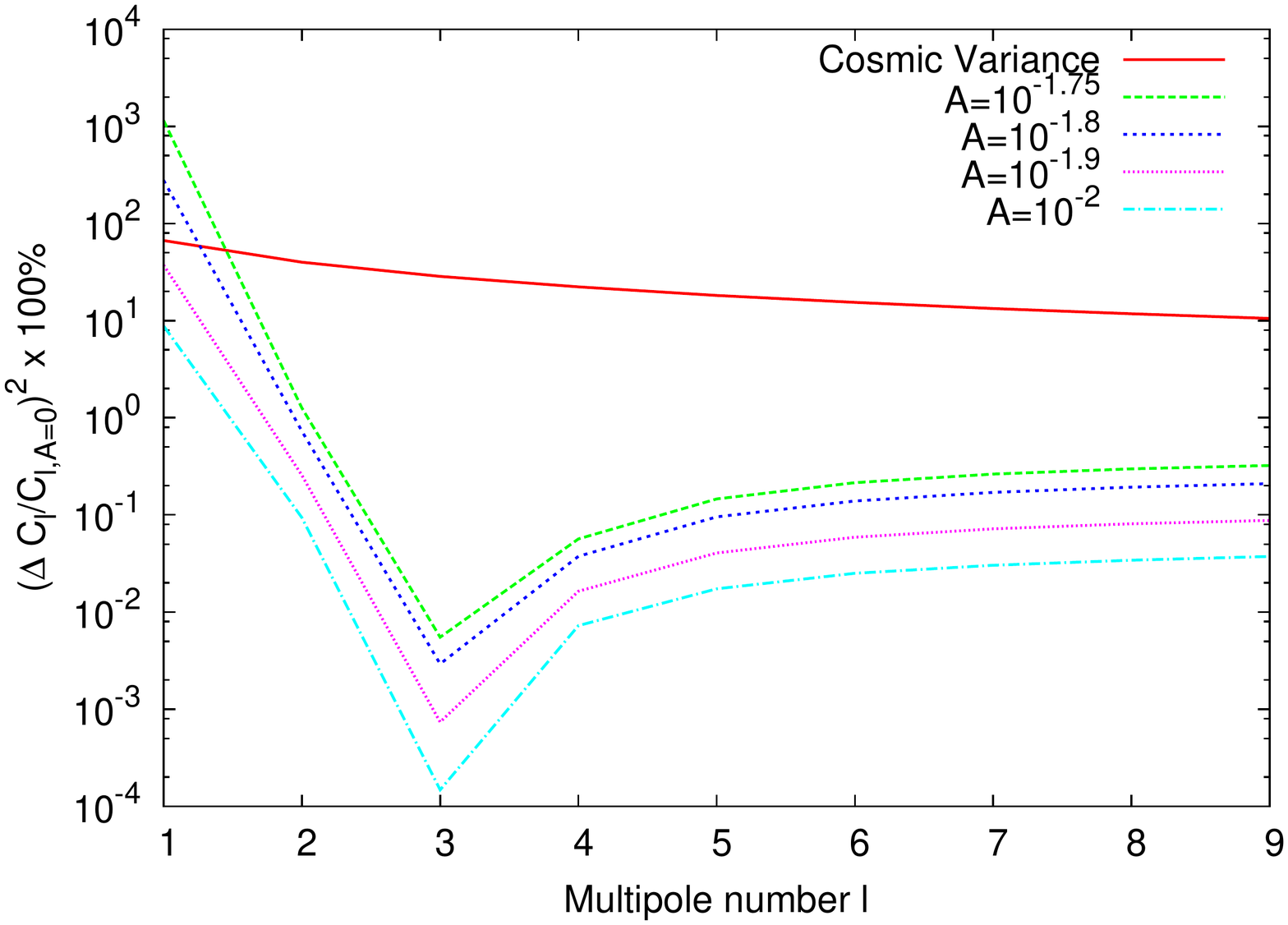}
	\caption{The fractional difference in the ISW effect in an interacting model relative to no interaction using the SUGRA potential.  Only the $\ell=1$ dipole moment of the largest two interaction strengths are observable above the cosmic variance.  The largest interaction strength leads to an order of magnitude enhancement to the CMB dipole moment relative to $\Lambda$CDM.    }
	\label{fig:11}
\end{figure}


The scalar-tensor model exhibits the most interesting behaviour of the three interaction models considered in this work.  The sign of the interaction can be changed to represent the flow of energy from one dark component to the other since the interaction depends both on dark matter and dark energy.  If we choose $Q_{\phi}=A\rho_{m}\dot{\phi}$ where $A$ is negative, we encounter the same instability in the $x$ variable which we found in the dark energy decay model.  From Equation$\nolinebreak \,\ref{Q16}$, we can easily deduce whether a model with $Q_{\phi}\propto x^{n}$ would develop an instability or not. When $n<1$, the interaction term becomes increasingly negative as $x$ decreases, hence leading to an instability.  For models where $n>1$, however, the negative constant tends to 0 as $x$ tends to 0 and so an instability is avoided. In this model $n=1$ and to avoid an instability we require that the sign of the interaction term represent flow of energy from dark matter to dark energy, i.e. $Q_{\phi}=A\rho_{m}\dot{\phi}$ with $A>0$. This supports the conclusion in \cite{2009PhRvD..79d3522C} which found no background instabilities in this model.  

We find that this form of interaction does not change the energy density of dark energy significantly at early times since $Q_{\phi} \propto \dot{\phi}$ and $\dot{\phi}$ is small at early times. The energy transfer becomes significant only at late times when dark energy becomes dominant.  As we increase the interaction strength we see an enhancement in the matter power spectrum on large scales, as shown in Figure$\nolinebreak \,\ref{fig:10}$. The power spectrum on smaller scales $k>0.01\mathrm{Mpc}^{-1}$ decreases by $\sim 6\%$ relative to $\Lambda$CDM for the largest interaction strength shown.

We have shown that a scalar-tensor interaction can lead to an enhancement in the matter power spectrum on large scales.  In order to detect this intrinsic enhancement we must be able to distinguish it from other artificial and observational effects such as gauge artifacts, the scale dependence of the galaxy bias and redshift space distortions which can also cause enhancements to the matter power spectrum on large scales and mimic our result (see \cite{2009PhRvD..80h3514Y}, \cite{Bruni:2011ta}, \cite{Challinor:2011bk} and \cite{Bonvin:2011bg}).  We will now address these effects in more detail and explain how to break the degeneracy between an intrinsic large-scale enhancement due to a scalar-tensor interaction and a similar enhancement caused by the effects above.

The work by \cite{2009PhRvD..80h3514Y} showed that calculating matter power spectra using gauge-dependent quantities can introduce gauge modes which can cause artificial large-scale enhancement in the power spectrum.  We have used gauge-invariant quantities in our calculations and so the large-scale enhancement we have found is not due to such a gauge artifact.  The linear bias is a linear relation between fractional overdensities of galaxies and matter ($\delta_{g}\propto\delta_{m}$) which is used to infer the total matter power spectrum from a galaxy survey.  Work by \cite{Bruni:2011ta} showed that the linear bias relation is scale-independent only in the comoving-synchronous gauge and using perturbations defined in a different gauge leads to the linear bias having a scale-dependence.  They showed that if the linear bias is used as a scale-independent relation in a different gauge it can lead to an artificial large-scale enhancement in the matter power spectrum \cite{Bruni:2011ta}.  Failure to use the correct form of the linear bias relation in a particular gauge could artificially replicate the signal from a scalar-tensor interaction.  

Redshift space distortions are caused by the peculiar velocities of observed luminous matter.  These small peculiar velocities produce an additional doppler shift in the observed photon frequency relative to the average redshift caused by the background expansion velocity of the region.  For large-scale overdensities the surrounding infalling matter will have an additional component of their peculiar velocities directed towards the overdensity due to gravitational attraction.  This leads to the region appearing compacted in redshift space, which causes an apparent enhancement to these large-scale overdensities (since redshift is used as a proxy for distance).  

Recent work by \cite{2009PhRvD..80h3514Y}, \cite{Challinor:2011bk} and \cite{Bonvin:2011bg} calculated the effects of redshift space distortions, magnification by gravitational lensing and distortions to the luminosity distances of sources on measured quantities such as the matter power spectrum.  They calculate photon geodesics in a perturbed universe using linear perturbation theory.  These geodesics take into account the deflection and redshifting of observed photons through interactions with overdensities.  They then use the perturbed geodesics of the observed photons to relate gauge-invariant observables from galaxy surveys (like the matter power spectrum) to the cosmological quantities calculated in theoretical models such as ours.  Using the results of \cite{2009PhRvD..80h3514Y}, \cite{Bruni:2011ta}, \cite{Challinor:2011bk} and \cite{Bonvin:2011bg} we can break the degeneracy between an intrinsic large-scale enhancement due to interactions in the dark sector and enhancements due to the artificial and observational effects discussed above.

Recently \cite{2011arXiv1103.0694T} investigated the effect of an interaction of the form ($\ref{Q15}$) on the background evolution, matter power spectrum and halo mass function using quintessence models.  They found that significant negative values of the interaction strength $A$ could lead to an enhancement in the matter power spectrum relative to $\Lambda$CDM, in agreement with our findings. The authors did not use fully gauge-invariant perturbation variables or include the effects of momentum transfer in their investigation, although they did include the effect of baryons and non-linearities. They did not find the instability in the $x$ variable when $A<0$, corresponding to dark energy decaying to dark matter. We expect that this is due to the substantially different treatment of the perturbations and the quintessence potentials they used. 

The production of large-scale enhancement of the matter power spectrum would be a very significant result if it could be resolved in future observations.  Recent work by \cite{2011MNRAS.tmp..391M} used peculiar velocities to constrain the matter power spectrum and found that an enhancement to the matter power spectrum on large scales was favoured by the data. However, due to large errors, primarily from cosmic variance, the findings were also consistent with $\Lambda$CDM.  Nevertheless, peculiar velocities could potentially be used to constrain the large scale matter power spectrum in the future when a greater number of accurate measurements are available.

Finally, Figure$\nolinebreak \,\ref{fig:11}$ shows how the ISW signal changes with increasing interaction strength.  We find that the interaction yields an enhancement in the ISW signal around only the lowest multipoles (since the dark energy spectrum peaks at small $k$ corresponding to small $\ell$). Only the $\ell=1$ moment of the two  strongest interaction strengths plotted ($A=10^{-1.75}\mpl^{-1}$ and $A=10^{-1.8}\mpl^{-1}$) is observable above cosmic variance.  The largest interaction strength leads to an order of magnitude enhancement to the CMB dipole moment relative to $\Lambda$CDM.

The recent high redshift survey by \cite{2010arXiv1012.2272T} found evidence of an excess clustering on large scales ($k<0.01h \mathrm{Mpc}^{-1}$) at 4$\sigma$ significance from $\Lambda$CDM.  Using photometric redshifts from a sample of SDSS galaxies they measured the angular power spectrum and found an excess in the lowest multipoles of the angular power spectrum.  This is a promising result, since it can be interpreted as the effect of the dark sector interaction shown in Figures $\ref{fig:10}$ and $\ref{fig:11}$.  After rigorous analysis of their data and biases they concluded that the large-scale enhancement was a real effect which hinted at an exotic form of dark energy.  

We find the recent evidence for enhanced power in $P(k)$ on large scales to be encouraging in light of our findings.  However, we can see from Figures $\ref{fig:10}$ and $\ref{fig:11}$ that even if there were significant increments in the matter power spectrum at these near-horizon scales, they are only detectable in the dipole moment of the CMB, a measurement which is dominated by the Doppler shift caused by our own peculiar velocity.  Unfortunately, the enhancements to the matter power spectrum on scales $k<0.001\mathrm{Mpc}^{-1}$ are well beyond the scales probed in current galaxy surveys ($k\sim0.01\mathrm{Mpc}^{-1}$), so we do not expect that a direct detection of this large scale enhancement via galaxy counts will be possible in the near future. 

Using the cross-correlation of the CMB and large-scale structure from galaxy surveys is likely to be one of the most effective observational probes of our results.  This technique allows us to isolate the ISW effect from other contributions to the CMB at low multipole moments since perturbations at the surface of last scattering were small compared to the observed large scale structure today.  Large scale structure is correlated with the ISW effect since the ISW effect is caused by the evolution of gravitational potentials.  Using the cross-correlation allows us to measure the ISW effect and compare this to the enhancement predicted in the scalar-tensor model for large interaction strengths (Figure $\ref{fig:11}$).  The work by \cite{Giannantonio:2008zi} calculated the cross-correlation using WMAP 7 and several galaxy surveys and found an excess ISW cross-correlation of 1$\sigma$ from $\Lambda$CDM.  This excess could be explained by a scalar-tensor dark sector interaction.


\section{Conclusion and discussion}

In this paper, we have calculated the background evolution, matter and dark energy power spectra and ISW effect for three quintessence models with and without dark sector interactions. Our calculations are based on gauge-invariant perturbation theory originally developed to calculate inflationary perturbations. We feel that our calculations are more reliable than those obtained when dynamical dark energy is parameterised by phenomenological ansatz for $w$, since in our approach the clustering of dark energy can be directly linked to the underlying scalar-field perturbations. 


We found that without interactions our quintessence models differ from $\Lambda$CDM by at most a few percent in observable quantities, meaning that any differences are essentially unobservable.  

We have shown how a covariant treatment of dark sector interactions can give rise to momentum exchanges in the dark sector, which many authors have  overlooked.  Our approach complements and extends the work of \cite{2008JCAP...07..020V}, who examined momentum exchanges in constant $w$ models using the dark matter decay interaction. We demonstrate our techniques on three model of interactions, namely, the dark energy decay, dark matter decay and scalar-tensor type of interaction.

The dark energy decay model was found to have an instability in the background and perturbation equations, in agreement with previous works on this interaction, and thus can be ruled out. In the decaying dark matter model, we found that its primary effect was to change the evolution of the background energy densities, with no observable differences in the power spectra or ISW effect.  We explained why  $w=$constant models, when associated with this interaction, will lead to an instability, which is absent when field dynamics are taken into account. 

The scalar-tensor model of interactions appears to be the most interesting and robust of all three. We showed that it is stable if energy flows from dark matter to dark energy.  For this case, we found that the interaction enhances the clustering of dark energy and dark matter on very large scales ($k<0.001\mathrm{Mpc}^{-1}$) and produced an enhanced ISW effect which exceeds cosmic variance only in the lowest multipoles.  We also found that the interaction produces a significant enhancement in the matter power spectra on very large scales.

We find the recent indications of enhanced large scale clustering via peculiar velocity measurements \cite{2011MNRAS.tmp..391M} and enhancement to the lowest multipoles of the angular power spectrum \cite{2010arXiv1012.2272T} encouraging, although, measurements on these scales are very difficult to make and are often dominated by cosmic variance and our own peculiar velocity.

Overall, the prospects of constraining dark sector interactions via the matter power spectrum and ISW effect appear daunting.  The enhancement to the matter power spectrum on scales $k<0.001\mathrm{Mpc}^{-1}$ is too large to be probed by current galaxy surveys $k\sim0.01\mathrm{Mpc}^{-1}$.  There are certainly other observational probes, however, which may reveal the presence of interactions in the dark sector, including i) cross-correlation of the ISW with galaxy and quasar distributions  \cite{Giannantonio:2008iv, Mainini:2010ng, Giannantonio:2008zi}, ii) the growth rates of large-scale structures at high redshift, \cite{Simpson:2010vh, 2010MNRAS.405.1006O, 2011arXiv1105.1194K}, iii) effects of large super-horizon perturbations on the CMB \cite{1978SvA....22..125G, 2011MNRAS.tmp..649B}, iv) gravitational lensing of the CMB using EPIC \cite{Martinelli:2010rt}. We envisage that our calculation techniques can be adapted to explore these issues at least in the linear regime.

Finally, there will almost certainly be interesting non-linear effects on the matter power spectrum and CMB anisotropies arising from dark sector interactions. However, analytic progress in this regime is extremely challenging (though not impossible \cite{Bernardeau:2001qr}). $N$-body simulations such as those initiated by \cite{Rasera:2010ar} may hold the key to understanding the non-linear effects of dark sector interactions.
 


\bibliography{Darkenergy}

\end{document}